\documentclass[aps,prl,twocolumn,nopacs,superscriptaddress]{revtex4}

\usepackage{graphicx}
\usepackage{verbatim}
\usepackage{relsize}
\usepackage{mathrsfs}
\usepackage{color}
\usepackage{mathtools}
\usepackage{array}
\usepackage{amsmath,amsfonts,amssymb}
\usepackage{graphicx}

\def\3{2.8in}    %used for figure widths
\def\2{2.5in}
\def\4{3.0in}

\def \beq {\begin{equation}}
\def \eeq {\end{equation}}
\begin{document}

\title{Topological crystalline insulator states in the Ca$_2$As family}
\author{Xiaoting Zhou$^*$}
\affiliation{Department of Physics, National Cheng Kung University, Tainan, 701, Taiwan}
\author{Chuang-Han Hsu$^*$}
\affiliation{Department of Physics, National University of Singapore, Singapore 117542}
\affiliation{Centre for Advanced 2D Materials and Graphene Research Centre, National University of Singapore, Singapore 117546}

\author{Tay-Rong Chang\footnote{These authors contributed equally to this work.}$^{\dag}$}
\affiliation{Department of Physics, National Cheng Kung University, Tainan, 701, Taiwan}
\author{ Hung-Ju Tien}
\affiliation{Department of Physics, National Cheng Kung University, Tainan, 701, Taiwan}

\author{Qiong Ma}\affiliation{Department of Physics, Massachusetts Institute of Technology, Cambridge, Massachusetts 02139, USA}

\author{Pablo Jarillo-Herrero}\affiliation {Department of Physics, Massachusetts Institute of Technology, Cambridge, Massachusetts 02139, USA}
\author{Nuh Gedik}\affiliation {Department of Physics, Massachusetts Institute of Technology, Cambridge, Massachusetts 02139, USA}

\author{Arun Bansil}
\affiliation{Department of Physics, Northeastern University, Boston, Massachusetts 02115, USA}
\author{Vitor M. Pereira}
\affiliation{Department of Physics, National University of Singapore, Singapore 117542}
\affiliation{Centre for Advanced 2D Materials and Graphene Research Centre, National University of Singapore, Singapore 117546}

\author{Su-Yang Xu$^{\dag}$}\affiliation {Department of Physics, Massachusetts Institute of Technology, Cambridge, Massachusetts 02139, USA}
\author{Hsin Lin}
\affiliation{Institute of Physics, Academia Sinica, Taipei 11529, Taiwan}
\author{Liang Fu\footnote{Corresponding authors' emails: 
u32trc00@phys.ncku.edu.tw, suyangxu@mit.edu, liangfu@mit.edu}}
\affiliation {Department of Physics, Massachusetts Institute of Technology, Cambridge, Massachusetts 02139, USA}

\begin{abstract}
{Topological crystalline insulators (TCI) are insulating electronic phases of matter with nontrivial topology originating from crystalline symmetries.  Recent theoretical advances have provided powerful guidelines to search for TCIs in real materials. Using density functional theory, we identify a class of new TCI states in the tetragonal lattice of the Ca$_2$As material family. On both top and side surfaces, we observe topological surface states protected independently by rotational and mirror symmetries. We show that a particular lattice distortion can single out the newly proposed topological protection by the rotational symmetry. As a result, the Dirac points of the topological surface states are moved to generic locations in momentum space away from any high symmetry lines. Such topological surface states have not been seen before. Moreover, the other family members, including Ca$_2$Sb, Ca$_2$Bi and Sr$_2$Sb, feature different topological surface states due to their distinct topological invariants. We thus further propose topological phase transitions in the pseudo-binary systems such as (Ca$_{1-x}$Sr$_x$)$_2$As and Ca$_2$As$_x$Sb$_{1-x}$. Our work reveals rich and exotic TCI physics across the Ca$_2$As family of materials, and suggests the feasibility of materials database search methods to discover new TCIs.}
\end{abstract}

\maketitle

Finding new topological phases of matter is a research frontier of modern condensed matter physics and materials science. Following the discoveries of time-reversal symmetry-protected $\mathcal{Z}_2$ topological insulator states in both 2D and 3D \cite{hasan2010colloquium, Qi2011, bansil2016colloquium}, intensive efforts have been devoted to searching for new types of topological band insulators protected by crystal symmetries, or topological crystalline insulators (TCI) \cite{fu2011topological}. Although the large number of crystalline space group symmetries suggests the possibility of a wealth of TCI states, for a long time, the theoretically known TCIs were largely limited to crystals with mirror reflections\cite{teo2008surface, fu2011topological, hsieh2012topological, weng2014topological} and glide mirror reflections \cite{wieder2017wallpaper, wang2016hourglass}, such as the SnTe family of
IV-VI semiconductors \cite{tanaka2012experimental, dziawa2012topological, xu2012TCI, okada2013observation, zeljkovic2014mapping, liang2013evidence, li2016interfacial, chang2016discovery, liang2017pressure}.

Recent work theorized a large new class of TCI states in time-reversal-invariant electron systems with spin-orbit coupling, which are protected by the $N$-fold rotational symmetries \cite{fang2017rotation}. Such rotational symmetry-protected TCIs are predicted to show distinct protected boundary states: The surface normal to the rotational  axis hosts $N$ Dirac cones, while any strictly two-dimensional system with the same symmetry necessarily has 2N stable Dirac points. In the absence of additional symmetry constraints, these $N$ Dirac cones are related by the $N$-fold rotational symmetry, and can appear at any generic $k$ points in the surface Brillouin zone. In addition, the side surface parallel to the rotational axis is predicted to host $N$ one-dimensional (1D) helical edge states, connecting the top and bottom surfaces. The existence of such 1D states was theoretically established in other classes of TCIs as well \cite{schindler2017higher, song2017d, matsugatani2018connecting}

\begin{figure*}
\includegraphics[width=165mm]{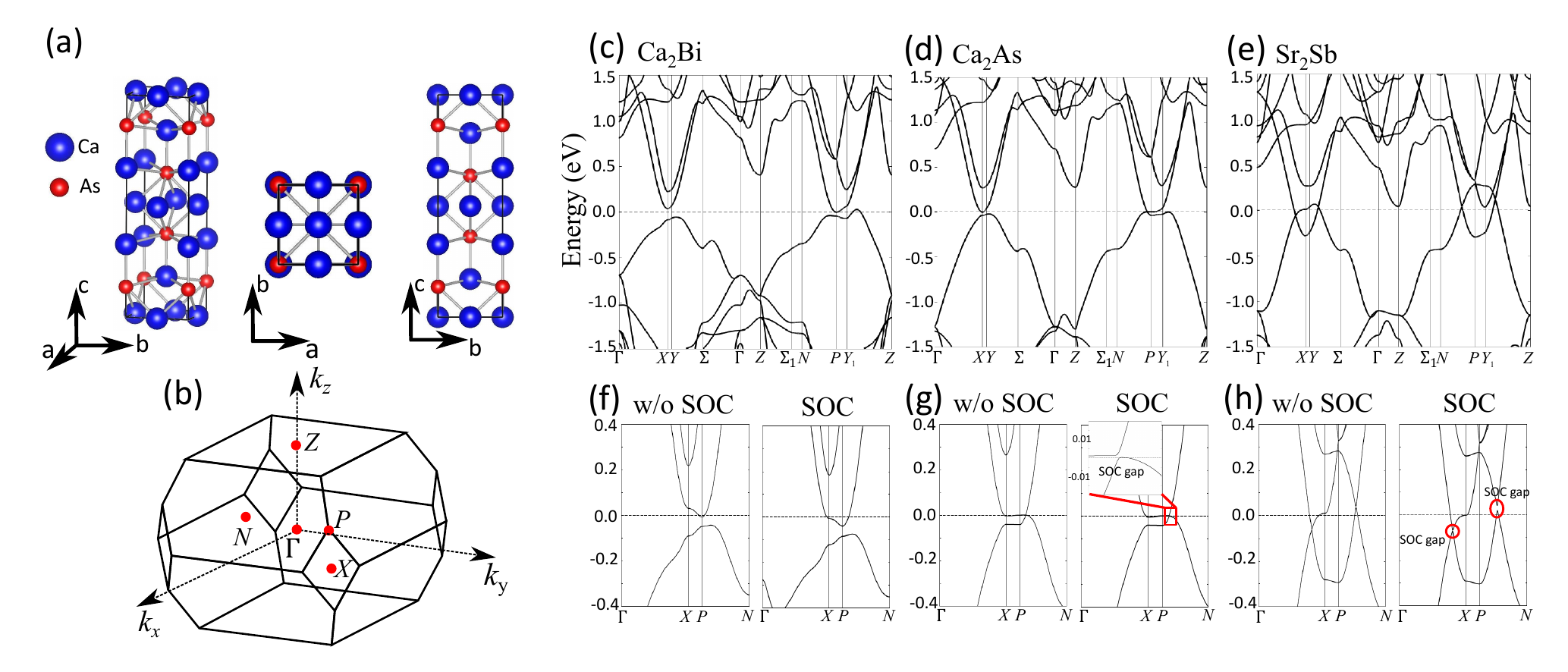}
\caption{\textbf{Lattice and electronic structure of the Ca$_2$As family}. (a) Crystal structure of Ca$_2$As's body centered tetragonal lattice. (b) The corresponding bulk Brillouin zone (BZ). (c-e) Band structures without spin-orbit coupling (SOC) over a wide energy window. (f-h) Zoomed-in band structures both without and with SOC.}
\label{Fig1}
\end{figure*}

\begin{figure}
\includegraphics[width=82mm]{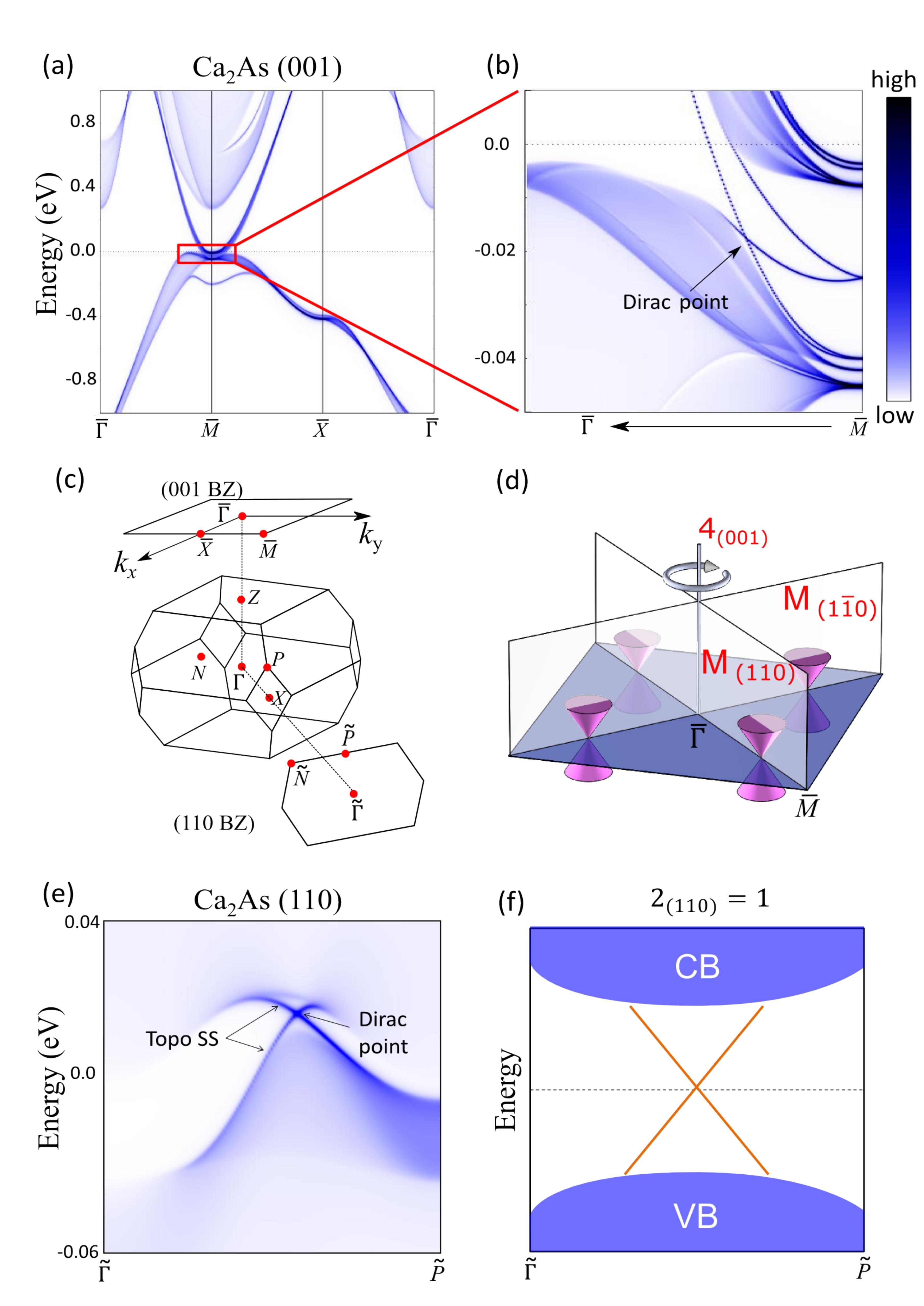}
\caption{\textbf{Topological surface states of Ca$_2$As on the $(001)$ and $(110)$ surfaces}. (a,b) Calculated surface spectral weight of the $(001)$ surface. The topological surface states along the $\bar{\Gamma}-\bar{M}$ line are denoted. (c) The $(001)$ and $(110)$ surface BZs. (d) Schematic illustration for the $4_{001}$ and $\mathcal{M}_{(110)}$($\mathcal{M}_{(1\bar{1}0)}$) topological protection of the surface states. (e) Calculated surface spectral weight of the $(110)$ surface. (f) Schematic illustration for the $2_{110}$ and $\mathcal{M}_{(1\bar{1}0)}$ topological protection of the surface states.}
\label{Fig2}
\end{figure}

\begin{figure*}
\includegraphics[width=150mm]{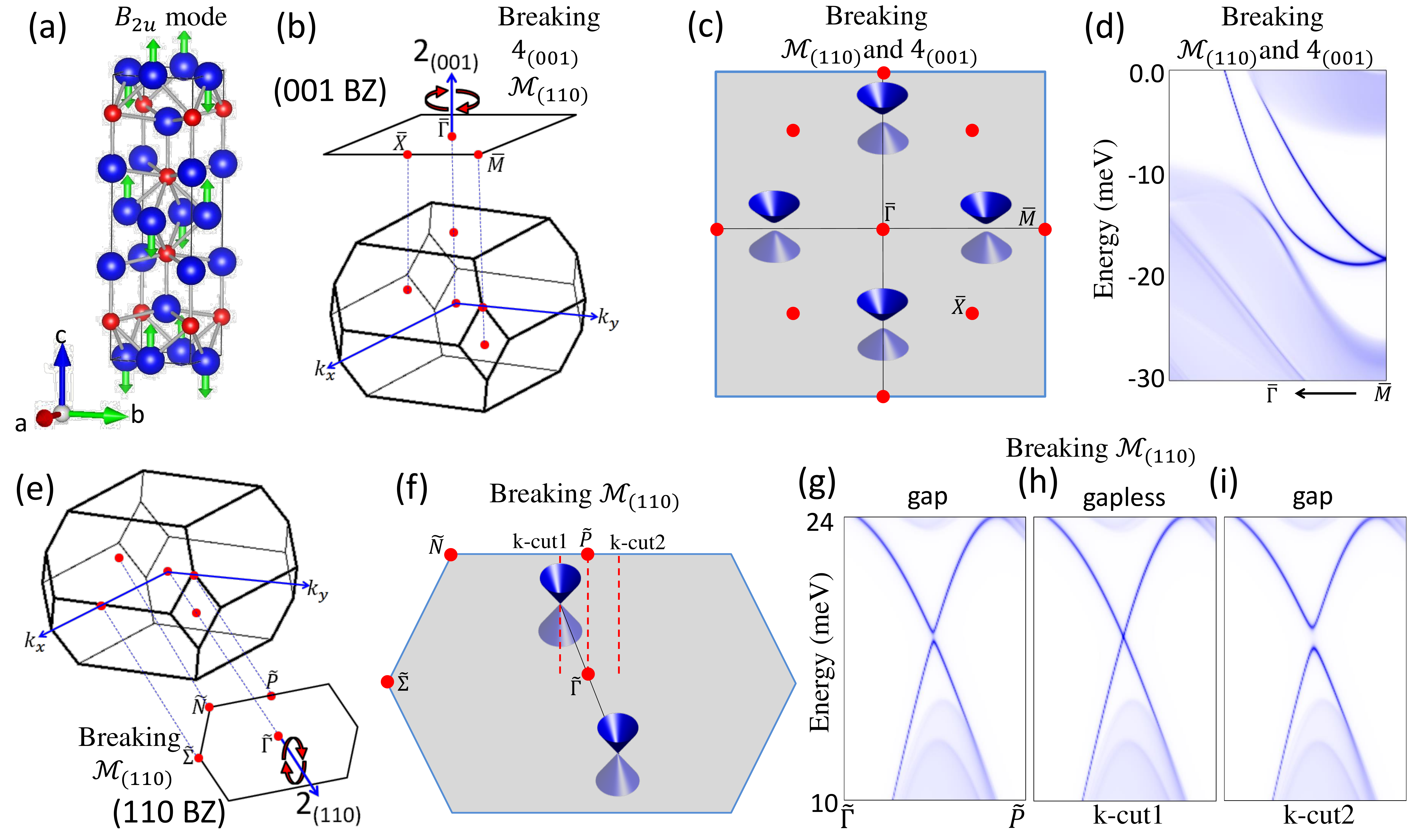}
\caption{\textbf{Isolating the rotational symmetry topological protection by lattice distortion}. (a) Atomic displacements according to the the $B_{2u}$ distortion. (b) Bulk and $(001)$ top surface BZs after the $B_{2u}$ distortion. (c) Because the $\mathcal{M}_{(110)}$, $\mathcal{M}_{(1\bar{1}0)}$, and $4_{(001)}$ symmetries are broken, all surface Dirac fermions on the $(001)$ surface are expected to gap out. (d) Calculated $(001)$ surface electronic structure after the $B_{2u}$ distortion shows the gap opening of the Dirac fermions. (e) Bulk and $(110)$  BZs after the $B_{2u}$ distortion. (f) Because $\mathcal{M}_{(110)}$, $\mathcal{M}_{(1\bar{1}0)}$ are broken but $2_{(110)}$ is preserved, the surface Dirac fermions on the $(110)$ side surface are expected to remain gapless but move away from the $\tilde{P}-\tilde{\Gamma}-\tilde{P}$ high symmetry line to generic $k$ points. (g-i) Calculated $(110)$ surface electronic structure after the $B_{2u}$ distortion shows that the Dirac fermions are indeed moved to away from generic $k$ points.}
\label{Fig3}
\end{figure*}

\begin{figure}
\includegraphics[width=88mm]{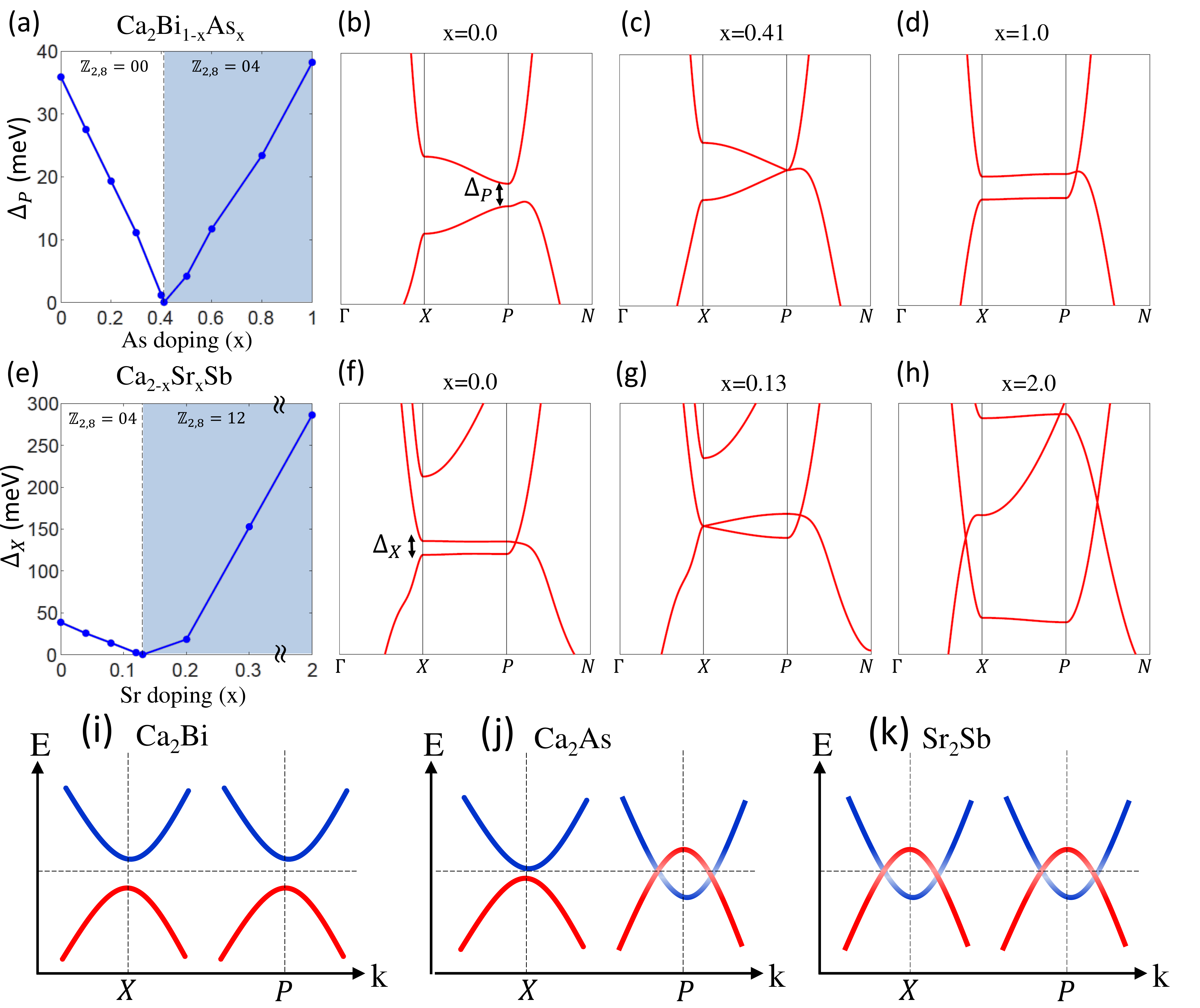}
\caption{\textbf{Topological phase transitions in Ca$_2$Bi$_{1-x}$As$_x$ and Ca$_{2-x}$Sr$_x$Sb}. (a) Calculated energy gap at the $P$ point in Ca$_2$Bi$_{1-x}$As$_x$. (b-d) Band structures at different $x$ values. (e) Calculated energy gap at the $X$ point in Ca$_{2-x}$Sr$_x$Sb. (f-h) Band structures at different $x$ values. (i-k) Schematic electronic states for the representative compounds in the Ca$_2$As family.}
\label{Fig4}
\end{figure}

The ubiquitous presence of rotational symmetries in crystals raises the hope that rotational symmetry-protected TCIs are widespread in real materials. However, their topological invariants defined in terms of Wannier center flow \cite{taherinejad2014wannier} are diffcult to compute in first-principles calculations. More generally, efficient and systematic methods to identify any class of TCIs from band structure calculations have been lacking - until very recently. Remarkably, Song et al \cite{song2017mapping} and Khalaf et al \cite{, khalaf2017symmetry}, building on earlier works \cite{bradlyn2017topological, po2017symmetry}, found that when certain additional symmetry $Y$ is present, topological invariants of TCIs protected by symmetry $X$ can be inferred by the $Y$-symmetry eigenvalues of energy bands. Such ``symmetry indicator'' of nontrivial topology is analogous to the parity criterion for time-reversal-invariant in the presence of inversion symmetry \cite{fu2007topological}. This band symmetry based approach bypasses the numerical difficulty in computing topological invariants directly, thus paving the way for systematic TCI materials search.

Taking advantage of these latest theoretical advances \cite{fang2017rotation, song2017d, bradlyn2017topological, po2017symmetry, song2017mapping, khalaf2017symmetry}, we propose a new type of TCI in the Ca$_2$As family materials, in which we identify topological surface states that are protected by both rotational symmetries and mirror symmetries. These novel topological electron states can lead to distinct signatures in various experiments including photoemission, scanning tunneling, transport, optical and optoelectronic  probes, which therefore pave the way for experimental studies on TCI physics.

\begin{table*}
\centering
\begin{tabular}{>{\centering\arraybackslash}m{4cm}>{\centering\arraybackslash}m{2cm}>{\centering\arraybackslash}m{2.5cm}>{\centering\arraybackslash}m{2cm}>{\centering\arraybackslash}m{2cm}>{\centering\arraybackslash}m{2cm}}
\hline
 & $(\mathcal{Z}_2\mathcal{Z}_8)$ & $(\nu_0;\nu_1\nu_2\nu_3)$ & $n_{\mathcal{M}_{(100)}}$ & $n_{\mathcal{M}_{(1\bar{1}0)}}$  & $n_{\mathcal{M}_{(001)}}$\\
\hline
$\textrm{Ca}_2\textrm{As and Ca}_2\textrm{Sb}$ & $(04)$ & $(0;000)$ & $0$ & $2$ & $0$\\ 
$\textrm{Sr}_2\textrm{Sb}$ & $(12)$ &$(0;111)$ & $0$ & $0$ & $2$\\ 
$\textrm{Ca}_2\textrm{Bi}$ & $(00)$ &$(0;000)$ & $0$ & $0$ & $0$\\ 
\hline
\end{tabular}
\caption{\textbf{Calculated symmetry indicator and topological invariants of the Ca$_2$Sb family.}}
\label{Tab1}
\end{table*}

The Ca$_2$As class of intermetallic materials crystallize in the body-centered tetragonal lattice system. The space group is $I4/mmm$ ($\#139$) and the point group is $D_{4h}$. The lattice is formed by a $\hat{z}$-direction stacking of alternating 2D square lattices of Ca or As atoms (Fig.~\ref{Fig1}(a)). The available symmetries include $\mathcal{I}$ (space inversion), $4_{(001)}$, $2_{(001)}$, $2_{(100)}$, $2_{(010)}$, $2_{(010)}$, $2_{(110)}$, $2_{(1\bar{1}0)}$, $\mathcal{M}_{(100)}$, $\mathcal{M}_{(010)}$, $\mathcal{M}_{(001)}$, $\mathcal{M}_{(110)}$, and $\mathcal{M}_{(1\bar{1}0)}$ \cite{notation}. Figure~\ref{Fig1}(b) shows the bulk Brillouin zone (BZ). The time-reversal invariant momenta (TRIM) include one $\Gamma$, two $X$, four $N$, and one $Z$, whereas other important non-TRIM high symmetry points are also noted. 

The low energy electronic structure of all family members is characterized by the two bands close to the Fermi energy (Figs.~\ref{Fig1}(c-e)). To examine the band inversion properties of each compound, we first study the band structure in the absence of spin-orbit coupling (SOC). In Ca$_2$As and Ca$_2$Sb (Fig.~\ref{Fig1}(g)), the lowest conduction band (CB) and the highest valence band (VB) are inverted near $P$ (non-TRIM) but remain non-inverted at other high symmetry points. In Sr$_2$Sb (Fig.~\ref{Fig1}(h)), the CB and VB are inverted in the vicinity of not only $P$ but also $X$ (TRIM). In Ca$_2$Bi (Fig.~\ref{Fig1}(f)), the CB and VB remain non-inverted everywhere in the BZ. The above band inversion properties suggest that Ca$_2$As and Ca$_2$Sb are pure TCIs, Sr$_2$Sb is both a weak TI (inverted at two TRIM $X$ points) and a TCI whereas Ca$_2$Bi is topologically trivial. 

\begin{table*}
\centering
\begin{tabular}{>{\centering\arraybackslash}m{2cm}>{\centering\arraybackslash}m{1.5cm}>{\centering\arraybackslash}m{1.5cm}>{\centering\arraybackslash}m{1.5cm}>{\centering\arraybackslash}m{1.5cm}>{\centering\arraybackslash}m{1.5cm}>{\centering\arraybackslash}m{1.5cm}>{\centering\arraybackslash}m{1.5cm}>{\centering\arraybackslash}m{1.5cm}>{\centering\arraybackslash}m{1.5cm}>{\centering\arraybackslash}m{1.5cm}}
\hline
 & $(\nu_0;\nu_1\nu_2\nu_3)$ & $n_{4_{(001)}}$ & $n_{2_{(100)}}$  & $n_{2_{(001)}}$ & $n_{2_{(110)}}$ & $n_{\mathcal{M}_{(1\bar{1}0)}}$  & $n_{\mathcal{M}_{(001)}}$ & $n_{\mathcal{M}_{(100)}}$ & $\mathcal{I}$\\
\hline
$\textrm{Ca}_2\textrm{As}$ & (0;000) & $1$ & $0$  & $0$ & $1$ & $2$  & $0$ & $0$ & $1$\\
\hline
$\textrm{Sr}_2\textrm{Sb}$ & (0;111) & $1$ & $1$  & $1$ & $1$ & $0$  & $2$ & $0$ & $1$\\
\hline
$\textrm{Ca}_2\textrm{Bi}$ & (0;000) & $0$ & $0$  & $0$ & $0$ & $0$  & $0$ & $0$ & $0$\\
\hline
\end{tabular}
\caption{\textbf{Calculated topological invariants of the Ca$_2$Sb family.}}
\label{Tab2}
\end{table*}

We now study the topology of these compounds by calculating the symmetry-based indicators. According to Ref. \cite{song2017mapping}, band insulators in space group $\#139$ are characterized by a pair of indicators $(\mathcal{Z}_2, \mathcal{Z}_8)$. By enumerating of valence electron states in certain irreducible presentations at specific high symmetry points (according to table I in Ref. \cite{song2017mapping}), we obtain the following $(\mathcal{Z}_2, \mathcal{Z}_8)$ values for indicators in table~\ref{Tab1}. Importantly, each value of symmetry indicators can point to several different possible topological states. Consulting table IV in Ref. \cite{song2017mapping}, we see that $(\mathcal{Z}_2, \mathcal{Z}_8)=(0,4)$ corresponds to four different pure TCI states; These four TCI states have different mirror Chern numbers $n_{\mathcal{M}_{(001)}}$ and $n_{\mathcal{M}_{(100)}}$.  $(\mathcal{Z}_2, \mathcal{Z}_8)=(1,2)$ corresponds to four ``weak TIs $+$ TCI'' states; Again, these four ``weak TIs $+$ TCI'' states have different mirror Chern numbers $n_{\mathcal{M}_{(001)}}$ and $n_{\mathcal{M}_{(100)}}$. $(\mathcal{Z}_2, \mathcal{Z}_8)=(0,0)$ corresponds to four states that are either completely trivial or a pure TCI. 

To nail down topological invariants uniquely, we now calculate the $\mathcal{Z}_2$ invariants $(\nu_0;\nu_1\nu_2\nu_3)$, the mirror Chern numbers $n_{\mathcal{M}_{(100)}}$ and $n_{\mathcal{M}_{(1\bar{1}0)}}$, and $n_{\mathcal{M}_{(001)}}$ (table~\ref{Tab1}). Through this way, we uniquely determine each compound's topological state (Table II) from the four possibilities inferred by the $(\mathcal{Z}_2, \mathcal{Z}_8)$ symmetry indicator. In the main text, we will focus on Ca$_2$As. We see from table~\ref{Tab2} that Ca$_2$As is a pure TCI: It features nontrivial rotational symmetry topological invariants ($n_{4_{(001)}}=1$ and $n_{2_{(110)}}=1$) in addition to the nontrivial mirror Chern number $n_{\mathcal{M}_{(1\bar{1}0)}}=2$.

In order to confirm these bulk topological invariants, we calculate the surface electronic structure of Ca$_2$As. We first focus on the $(001)$ surface. Since the bulk band inversion occurs near the $P$ point whose surface projection is $\bar{M}$ (Fig.~\ref{Fig2}(c)), we expect the low energy electronic structure of the $(001)$ surface band structure to be in the vicinity of $\bar{M}$. Figure.~\ref{Fig2}(a) indeed confirms this case. Zooming in near the $\bar{M}$ point, we observe (Fig.~\ref{Fig2}(b)) clear surface states (sharp lines) dispersing inside the projected bulk energy gap. Specifically, the surface states form a Dirac point along the $\bar{\Gamma}-\bar{M}$ direction in the vicinity of the $\bar{M}$ point. In Fig.~\ref{Fig2}(d), we show an overview of the topological surface states in the surface BZ: there are four topological Dirac surface states located along the $\bar{M}-\bar{\Gamma}-\bar{M}$ lines. We check the consistency between the bulk topological invariants and the topological surface states: On the $(001)$ surface, the four-fold rotation ($4_{(001)}$) and the mirror planes ($\mathcal{M}_{(110)}$ and $\mathcal{M}_{(1\bar{1}0)}$) are projected as the $\bar{\Gamma}$ point and the $\bar{M}-\bar{\Gamma}-\bar{M}$ lines, respectively. By contrast, the two-fold rotation $2_{(110)}$ is not respected and thus becomes not important for the $(001)$ surface. Consulting table~\ref{Tab2}, the nontrivial mirror Chern number ($n_{\mathcal{M}_{(110)}}=n_{\mathcal{M}_{(1\bar{1}0)}}=2$) requires two Dirac surface states on each mirror line; the nontrivial rotational topological invariant ($n_{4_{(001)}}=1$) requires four Dirac surface states that are related by the rotational center. Therefore, the obtained surface states are consistent with the bulk topological invariants. On the other hand, the $(110)$ surface shows two surface Dirac cones along the $\tilde{P}-\tilde{\Gamma}-\tilde{P}$ direction (Figs.~\ref{Fig2}(e-f)), which are consistent with the topological invariants $n_{\mathcal{M}_{(1\bar{1}0)}}=2$ and $n_{2_{(110)}}=1$. These predicted topological surface states can be detected by photoemission experiments on the Ca$_2$As family of compounds.

We now show that a particular kind of lattice distortion can isolate the topological protection by rotational symmetry. The strategy is to identify a particular lattice distortion that can break mirror symmetries while keeping rotational symmetries intact; The rotation-protected topology should remain robust as long as the distortion does not affect the band inversion. As a guideline, we calculated the phonon spectrum and identified phonons in the $A_{1g}$, $E_g$, $A_{2u}$, $B_{2u}$, and $E_u$  presentations at the $\Gamma$ point. Our symmetry analysis shows that a distortion according to the $B_{2u}$ modes can break $\mathcal{M}_{(110)}$, $\mathcal{M}_{(1\bar{1}0)}$, and $4_{(001)}$ but preserve $2_{(110)}$ (Fig.~\ref{Fig3}(a)). The resulting space group is $I-4m2 \ (\#119)$. Figure~\ref{Fig3} shows the surface band structure after a $B_{2u}$ distortion. Remarkably, we found that all surface Dirac cones on the top $(001)$ surface are gapped (Figs.~\ref{Fig3}(b-d)); By contrast, the Dirac cones on the side $(110)$ surface remain gapless but their Dirac points are moved to generic $k$ points off the $\tilde{P}-\tilde{\Gamma}-\tilde{P}$ line (Figs.~\ref{Fig3}(e-i)). These Dirac cones thus represent a new type of topological surface states that are solely protected by the rotational symmetries. The odd parity phonon can be directly driven by terahertz optical pulse \cite{forst2015mode}. Theory also predicted the existence of 1D helical edge states along the side surfaces, which require extremely large computational cost. This can be a direction for future works. 

\color{black} We propose topological phase transitions in the pseudo-binary systems. As shown in Fig.~\ref{Fig4}(a), substituting As for Bi in Ca$_2$Bi leads to a topological phase transition from a trivial insulator (Ca$_2$Bi, ($\mathcal{Z}_2\mathcal{Z}_8)=(00)$) to a pure TCI (Ca$_2$As, ($\mathcal{Z}_2\mathcal{Z}_8)=(04)$). The critical point lies at the doping level $x\simeq41\%$ (Figs.~\ref{Fig4}(a-c)). Through this transition, we expect topological Dirac surface states to emerge at the $(001)$ and $(110)$ surfaces. Similarly, we predict the Ca$_{2-x}$Sr$_x$Sb system to realize a topological phase transition from a pure TCI (Ca$_2$Sb, ($\mathcal{Z}_2\mathcal{Z}_8)=(04)$) to a TCI/weak TI (Sr$_2$Sb, ($\mathcal{Z}_2\mathcal{Z}_8)=(12)$). We found that the critical Sr doping level is roughly at $x\simeq13\%$ (Figs.~\ref{Fig4}(e-g)). We note that the random substitutions in the pseudo-binary systems do not preserve the local crystalline symmetries. Therefore, proposed pseudo-binary systems also provide a platform to study the robustness of the nontrivial topology against average crystalline symmetries \cite{fu2012topology}.

\textbf{Acknowledgements}
We thank Chen Fang for valuable discussions on the symmetry indictors. L.F. was supported by the US Department of Energy (DOE), Office of Science, Office of Basic Energy Sciences (BES), Division of Materials Sciences and Engineering (DMSE) under award DE-SC0010526. Q.M. and P.J.H. were supported by the Center for Excitonics, an Energy Frontier Research Center funded by the DOE, Office of Science, BES under Award Number DESC0001088 and AFOSR grant FA9550-16-1-0382, as well as the Gordon and Betty Moore Foundation's EPiQS Initiative through Grant GBMF4541 to P.J.H.. N.G. and S.Y.X. acknowledge support from DOE, BES DMSE and the Gordon and Betty Moore Foundations EPiQS Initiative through Grant GBMF4540. T.-R.C., X.Z. and H.-J.T. were supported by the Ministry of Science and Technology under MOST Young Scholar Fellowship: the MOST Grant for the Columbus Program NO. 107-2636-M-006 -004 -, National Cheng Kung University, Taiwan, and National Center for Theoretical Sciences (NCTS), Taiwan.

\color{black}
\vspace{1cm}

\bibliography{Topological_and_2D_05102018}

\begin{thebibliography}{10}
\expandafter\ifx\csname url\endcsname\relax
  \def\url#1{\texttt{#1}}\fi
\expandafter\ifx\csname urlprefix\endcsname\relax\def\urlprefix{URL }\fi
\providecommand{\bibinfo}[2]{#2}
\providecommand{\eprint}[2][]{\url{#2}}

\bibitem{hasan2010colloquium}
\bibinfo{author}{Hasan, M.~Z.} \& \bibinfo{author}{Kane, C.~L.}
\newblock \bibinfo{title}{Colloquium: {T}opological {I}nsulators}.
\newblock \emph{\bibinfo{journal}{Reviews of Modern Physics}}
  \textbf{\bibinfo{volume}{82}}, \bibinfo{pages}{3045--3067}
  (\bibinfo{year}{2010}).

\bibitem{Qi2011}
\bibinfo{author}{Qi, X.-L.} \& \bibinfo{author}{Zhang, S.-C.}
\newblock \bibinfo{title}{Topological insulators and superconductors}.
\newblock \emph{\bibinfo{journal}{Rev. Mod. Phys.}}
  \textbf{\bibinfo{volume}{83}}, \bibinfo{pages}{1057--1110}
  (\bibinfo{year}{2011}).
\newblock \urlprefix\url{http://link.aps.org/doi/10.1103/RevModPhys.83.1057}.

\bibitem{bansil2016colloquium}
\bibinfo{author}{Bansil, A.}, \bibinfo{author}{Lin, H.} \&
  \bibinfo{author}{Das, T.}
\newblock \bibinfo{title}{Colloquium: Topological band theory}.
\newblock \emph{\bibinfo{journal}{Reviews of Modern Physics}}
  \textbf{\bibinfo{volume}{88}}, \bibinfo{pages}{021004}
  (\bibinfo{year}{2016}).

\bibitem{fu2011topological}
\bibinfo{author}{Fu, L.}
\newblock \bibinfo{title}{Topological crystalline insulators}.
\newblock \emph{\bibinfo{journal}{Phys. Rev. Lett.}}
  \textbf{\bibinfo{volume}{106}}, \bibinfo{pages}{106802}
  (\bibinfo{year}{2011}).

\bibitem{teo2008surface}
\bibinfo{author}{Teo, J.~C.}, \bibinfo{author}{Fu, L.} \&
  \bibinfo{author}{Kane, C.}
\newblock \bibinfo{title}{Surface states and topological invariants in
  three-dimensional topological insulators: {A}pplication to {B}i$_{1- x}$
  {S}b$_x$}.
\newblock \emph{\bibinfo{journal}{Physical Review B}}
  \textbf{\bibinfo{volume}{78}}, \bibinfo{pages}{045426}
  (\bibinfo{year}{2008}).

\bibitem{hsieh2012topological}
\bibinfo{author}{Hsieh, T.~H.} \emph{et~al.}
\newblock \bibinfo{title}{Topological crystalline insulators in the {S}n{T}e
  material class}.
\newblock \emph{\bibinfo{journal}{Nature Communications}}
  \textbf{\bibinfo{volume}{3}}, \bibinfo{pages}{982} (\bibinfo{year}{2012}).

\bibitem{weng2014topological}
\bibinfo{author}{Weng, H.}, \bibinfo{author}{Zhao, J.}, \bibinfo{author}{Wang,
  Z.}, \bibinfo{author}{Fang, Z.} \& \bibinfo{author}{Dai, X.}
\newblock \bibinfo{title}{Topological crystalline kondo insulator in mixed
  valence ytterbium borides}.
\newblock \emph{\bibinfo{journal}{Physical review letters}}
  \textbf{\bibinfo{volume}{112}}, \bibinfo{pages}{016403}
  (\bibinfo{year}{2014}).

\bibitem{wieder2017wallpaper}
\bibinfo{author}{Wieder, B.~J.} \emph{et~al.}
\newblock \bibinfo{title}{Wallpaper {F}ermions and the {T}opological {D}irac
  {I}nsulator}.
\newblock \emph{\bibinfo{journal}{arXiv preprint arXiv:1705.01617}}
  (\bibinfo{year}{2017}).

\bibitem{wang2016hourglass}
\bibinfo{author}{Wang, Z.}, \bibinfo{author}{Alexandradinata, A.},
  \bibinfo{author}{Cava, R.~J.} \& \bibinfo{author}{Bernevig, B.~A.}
\newblock \bibinfo{title}{Hourglass fermions}.
\newblock \emph{\bibinfo{journal}{Nature}} \textbf{\bibinfo{volume}{532}},
  \bibinfo{pages}{189--194} (\bibinfo{year}{2016}).

\bibitem{tanaka2012experimental}
\bibinfo{author}{Tanaka, Y.} \emph{et~al.}
\newblock \bibinfo{title}{Experimental realization of a topological crystalline
  insulator in {S}n{T}e}.
\newblock \emph{\bibinfo{journal}{Nature Physics}}
  \textbf{\bibinfo{volume}{8}}, \bibinfo{pages}{800--803}
  (\bibinfo{year}{2012}).

\bibitem{dziawa2012topological}
\bibinfo{author}{Dziawa, P.} \emph{et~al.}
\newblock \bibinfo{title}{Topological crystalline insulator states in
  {P}b$_{1-x}${S}n$_x${S}e}.
\newblock \emph{\bibinfo{journal}{Nature Materials}}
  \textbf{\bibinfo{volume}{11}}, \bibinfo{pages}{1023--1027}
  (\bibinfo{year}{2012}).

\bibitem{xu2012TCI}
\bibinfo{author}{Xu, S.-Y.} \emph{et~al.}
\newblock \bibinfo{title}{Observation of a topological crystalline insulator
  phase and topological phase transition in {P}b$_{1-x}${S}n$_x${T}e}.
\newblock \emph{\bibinfo{journal}{Nature Communications}}
  \textbf{\bibinfo{volume}{3}}, \bibinfo{pages}{1192} (\bibinfo{year}{2012}).

\bibitem{okada2013observation}
\bibinfo{author}{Okada, Y.} \emph{et~al.}
\newblock \bibinfo{title}{Observation of {D}irac node formation and mass
  acquisition in a topological crystalline insulator}.
\newblock \emph{\bibinfo{journal}{Science}} \textbf{\bibinfo{volume}{341}},
  \bibinfo{pages}{1496--1499} (\bibinfo{year}{2013}).

\bibitem{zeljkovic2014mapping}
\bibinfo{author}{Zeljkovic, I.} \emph{et~al.}
\newblock \bibinfo{title}{Mapping the unconventional orbital texture in
  topological crystalline insulators}.
\newblock \emph{\bibinfo{journal}{Nature Physics}}
  \textbf{\bibinfo{volume}{10}}, \bibinfo{pages}{572--577}
  (\bibinfo{year}{2014}).

\bibitem{liang2013evidence}
\bibinfo{author}{Liang, T.} \emph{et~al.}
\newblock \bibinfo{title}{Evidence for massive bulk {D}irac fermions in
  {P}b$_{1-x}${S}n$_x${S}e from {N}ernst and thermopower experiments.}
\newblock \emph{\bibinfo{journal}{Nature communications}}
  \textbf{\bibinfo{volume}{4}}, \bibinfo{pages}{2696} (\bibinfo{year}{2013}).

\bibitem{li2016interfacial}
\bibinfo{author}{Li, X.} \& \bibinfo{author}{Niu, Q.}
\newblock \bibinfo{title}{Interfacial-state coupling induced topological phase
  transitions in snte (110) thin film}.
\newblock \emph{\bibinfo{journal}{arXiv preprint arXiv:1611.03151}}
  (\bibinfo{year}{2016}).

\bibitem{chang2016discovery}
\bibinfo{author}{Chang, K.} \emph{et~al.}
\newblock \bibinfo{title}{Discovery of robust in-plane ferroelectricity in
  atomic-thick {S}n{T}e}.
\newblock \emph{\bibinfo{journal}{Science}} \textbf{\bibinfo{volume}{353}},
  \bibinfo{pages}{274--278} (\bibinfo{year}{2016}).

\bibitem{liang2017pressure}
\bibinfo{author}{Liang, T.} \emph{et~al.}
\newblock \bibinfo{title}{A pressure-induced topological phase with large
  {B}erry curvature in {P}b$_{1-x}${S}n$_x${T}e}.
\newblock \emph{\bibinfo{journal}{Science Advances}}
  \textbf{\bibinfo{volume}{3}}, \bibinfo{pages}{e1602510}
  (\bibinfo{year}{2017}).

\bibitem{fang2017rotation}
\bibinfo{author}{Fang, C.} \& \bibinfo{author}{Fu, L.}
\newblock \bibinfo{title}{Rotation {A}nomaly and {T}opological {C}rystalline
  {I}nsulators}.
\newblock \emph{\bibinfo{journal}{arXiv preprint arXiv:1709.01929}}
  (\bibinfo{year}{2017}).

\bibitem{schindler2017higher}
\bibinfo{author}{Schindler, F.} \emph{et~al.}
\newblock \bibinfo{title}{Higher-{O}rder {T}opological {I}nsulators}.
\newblock \emph{\bibinfo{journal}{arXiv preprint arXiv:1708.03636}}
  (\bibinfo{year}{2017}).

\bibitem{song2017d}
\bibinfo{author}{Song, Z.}, \bibinfo{author}{Fang, Z.} \&
  \bibinfo{author}{Fang, C.}
\newblock \bibinfo{title}{$(d-2)$-{D}imensional {E}dge {S}tates of {R}otation
  {S}ymmetry {P}rotected {T}opological {S}tates}.
\newblock \emph{\bibinfo{journal}{Phys. Rev. Lett.}}
  \textbf{\bibinfo{volume}{119}}, \bibinfo{pages}{246402}
  (\bibinfo{year}{2017}).

\bibitem{matsugatani2018connecting}
\bibinfo{author}{Matsugatani, A.} \& \bibinfo{author}{Watanabe, H.}
\newblock \bibinfo{title}{Connecting higher-order topological insulators to
  lower-dimensional topological insulators}.
\newblock \emph{\bibinfo{journal}{arXiv preprint arXiv:1804.02794}}
  (\bibinfo{year}{2018}).

\bibitem{taherinejad2014wannier}
\bibinfo{author}{Taherinejad, M.}, \bibinfo{author}{Garrity, K.~F.} \&
  \bibinfo{author}{Vanderbilt, D.}
\newblock \bibinfo{title}{Wannier center sheets in topological insulators}.
\newblock \emph{\bibinfo{journal}{Physical Review B}}
  \textbf{\bibinfo{volume}{89}}, \bibinfo{pages}{115102}
  (\bibinfo{year}{2014}).

\bibitem{song2017mapping}
\bibinfo{author}{Song, Z.}, \bibinfo{author}{Zhang, T.}, \bibinfo{author}{Fang,
  Z.} \& \bibinfo{author}{Fang, C.}
\newblock \bibinfo{title}{Mapping symmetry data to topological invariants in
  nonmagnetic materials}.
\newblock \emph{\bibinfo{journal}{arXiv preprint arXiv:1711.11049}}
  (\bibinfo{year}{2017}).

\bibitem{khalaf2017symmetry}
\bibinfo{author}{Khalaf, E.}, \bibinfo{author}{Po, H.~C.},
  \bibinfo{author}{Vishwanath, A.} \& \bibinfo{author}{Watanabe, H.}
\newblock \bibinfo{title}{Symmetry indicators and anomalous surface states of
  topological crystalline insulators}.
\newblock \emph{\bibinfo{journal}{arXiv preprint arXiv:1711.11589}}
  (\bibinfo{year}{2017}).

\bibitem{bradlyn2017topological}
\bibinfo{author}{Bradlyn, B.} \emph{et~al.}
\newblock \bibinfo{title}{Topological quantum chemistry}.
\newblock \emph{\bibinfo{journal}{Nature}} \textbf{\bibinfo{volume}{547}},
  \bibinfo{pages}{298} (\bibinfo{year}{2017}).

\bibitem{po2017symmetry}
\bibinfo{author}{Po, H.~C.}, \bibinfo{author}{Vishwanath, A.} \&
  \bibinfo{author}{Watanabe, H.}
\newblock \bibinfo{title}{Symmetry-based indicators of band topology in the
  $230$ space groups}.
\newblock \emph{\bibinfo{journal}{Nature {C}ommunications}}
  \textbf{\bibinfo{volume}{8}}, \bibinfo{pages}{50} (\bibinfo{year}{2017}).

\bibitem{fu2007topological}
\bibinfo{author}{Fu, L.} \& \bibinfo{author}{Kane, C.~L.}
\newblock \bibinfo{title}{Topological insulators with inversion symmetry}.
\newblock \emph{\bibinfo{journal}{Phys. Rev. B}} \textbf{\bibinfo{volume}{76}},
  \bibinfo{pages}{045302} (\bibinfo{year}{2007}).

\bibitem{notation}
\bibinfo{title}{In our notation, $2_{(100)}$ means a $2$-fold rotation about
  the (100) crystalline axis; $\mathcal{M}_{(100)}$ means a mirror operation
  that reflects the $(100)$ crystalline direction to $(\bar{1}00)$.}

\bibitem{forst2015mode}
\bibinfo{author}{F{\"o}rst, M.}, \bibinfo{author}{Mankowsky, R.} \&
  \bibinfo{author}{Cavalleri, A.}
\newblock \bibinfo{title}{Mode-selective control of the crystal lattice}.
\newblock \emph{\bibinfo{journal}{Accounts of chemical research}}
  \textbf{\bibinfo{volume}{48}}, \bibinfo{pages}{380--387}
  (\bibinfo{year}{2015}).

\bibitem{fu2012topology}
\bibinfo{author}{Fu, L.} \& \bibinfo{author}{Kane, C.~L.}
\newblock \bibinfo{title}{Topology, delocalization via average symmetry and the
  symplectic anderson transition}.
\newblock \emph{\bibinfo{journal}{Phys. Rev. Lett.}}
  \textbf{\bibinfo{volume}{109}}, \bibinfo{pages}{246605}
  (\bibinfo{year}{2012}).

\end{thebibliography}

\bibliographystyle{naturemag}

\end{document}